# Terahertz-driven, all-optical electron gun

W. Ronny Huang[1,2], Arya Fallahi[1], Xiaojun Wu[1], Huseyin Cankaya[1], Anne-Laure Calendron[1], Koustuban Ravi[1,2], Dongfang Zhang[1], Emilio A. Nanni[2], Kyung-Han Hong[2], Franz X. Kärtner[1,2,*]

[1]Center for Free-Electron Laser Science at Deutsches Electronen Synchrotron (DESY), Department of Physics and the Center for Ultrafast Imaging, University of Hamburg, Hamburg 22761, Germany

[2]Department of Electrical Engineering and Computer Science and Research Laboratory of Electronics, Massachusetts Institute of Technology (MIT), Cambridge, Massachusetts 02139, USA

*Corresponding author: franz.kaertner@cfel.de.

## 1. Introduction

**Ultrashort electron beams with narrow energy spread, high charge, and low jitter are essential for resolving phase transitions in metals [Siwick2003], semiconductors [Morrison2014], and molecular crystals [Ishikawa2015]. These semirelativistic beams, produced by phototriggered electron guns, are also injected into accelerators for x-ray light sources [Kaertner2016]. The achievable resolution of these time-resolved electron diffraction or x-ray experiments has been hindered by surface field and timing jitter limitations in conventional RF guns, which thus far are <200 MV/m [Zhou2010] and >96 fs [Brussaard2013], respectively. A gun driven by optically-generated single-cycle THz pulses provides a practical solution to enable not only GV/m surface fields [Loew1988, Dolgashev2010] but also absolute timing stability, since the pulses are generated by the same laser as the phototrigger. Here, we demonstrate an all-optical THz gun yielding peak electron energies approaching 1 keV, accelerated by >300 MV/m THz fields in a novel micron-scale waveguide structure. We also achieve quasimonoenergetic, sub-keV bunches with 32 fC of charge, which can already be used for time-resolved low-energy electron diffraction (LEED) [Gulde2014]. Such ultracompact, easy-to-implement guns—driven by intrinsically-synchronized THz pulses that are pumped by an amplified arm of the already-present photoinjector laser—provide a new tool with potential to transform accelerator-based science.**

The central challenge of an electron gun is to accelerate electrons from rest to relativistic energies as quickly as possible to avoid the beam-degrading effects of space charge, which scale inversely as the electron energy squared [Wiedemann2007], and hence, inversely as the accelerating field squared. To achieve the desired high fields, there are currently two types of electron guns, DC and RF guns, which have field limitations of around 10 MV/m [Loehl2010] and 200 MV/m [Zhou2010], respectively, due to breakdown mechanisms on common accelerator materials [Dolgashev2010,Forno2016,Laurent2011]. DC guns utilize high voltage electrodes, which require enormous power supplies and bulky feedthroughs. RF guns on the other hand utilize high power RF fields, which involve expensive klystrons, pulsed heating issues [Dolgashev2010,Laurent2011], and elaborate synchronization schemes [Brussaard2013,Harmand2012]. The need for a more compact, economical electron source with higher accelerating field, that may ultimately lead to lower emittance electron bunches [Engelen2013], has propelled the development of photonic (IR- or THz-driven) linear accelerators (linacs) with promising results [Peralta2013,Nanni2015]. However, the potential advantages of photonic linacs have not extended to *photonic guns*, the initial acceleration stage that is quintessential to determining the final electron beam quality. The difficulty, which lies primarily in phase matching the electromagnetic wave with nonrelativistic electrons, is greater for short IR wavelengths [Zawadzka2001] than for THz radiation. Thus, we recently proposed the development of a single-cycle THz gun [Huang2015,Fallahi2016] to exploit the GV/m fields possible with optically-generated THz sources [Shalaby2015]. Here, we implement such a THz gun. Leveraging the gun's simple geometries and flexible machining requirements, we integrate it in a practical, compact machine that is powered by a 1 kHz, few-mJ laser and operates without external synchronization. Our first results demonstrate high field (350 MV/m) THz acceleration up to 0.8 keV, as well as percent-level energy spread in sub-keV, multi-10 fC bunches. These results, which are already suitable for time-resolved LEED experiments, confirm the performance of a THz-driven gun technology that is scalable to relativistic energies [Fallahi2016].

## 2. Experimental setup

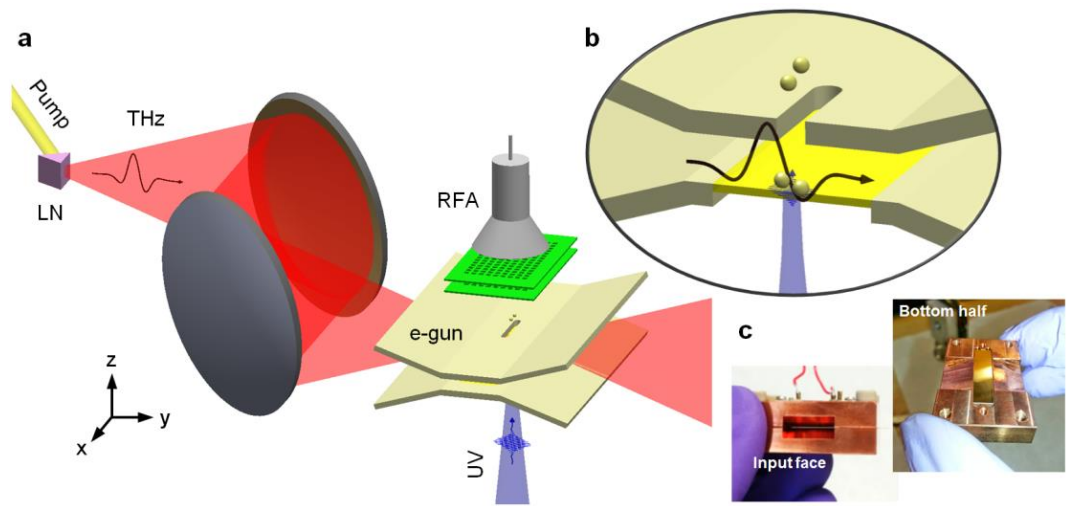

**Figure fig:schem | Schematic of the THz gun. (a) A single-cycle THz pulse, generated via optical rectification in lithium niobate (LN), is coupled into the THz gun, which takes the form of a parallel-plate waveguide (PPWG) for field confinement. A UV pulse backilluminates the photocathode to emit an electron bunch, which is subsequently accelerated by the THz field. The bunch exits through a slit in the top plate and a retarding field analyzer (RFA) measures its energy spectrum. (b) Cross section of the gun, showing the UV-photoemitted electrons being accelerated by the THz field and escaping through the slit. (c) Photographs of the THz gun. (Left) View facing the input taper of the PPWG. (Right) Bottom half of the PPWG with the photocathode in the middle.**

The THz gun (Figure fig:schem(a)-(c)), takes the form of a copper parallel-plate waveguide (PPWG) with a subwavelength spacing of 75 µm. We exploit this structure's TEM mode for unchirped, uniform enhancement of the THz field [Iwaszczuk2012]. A freespace z-polarized THz beam is coupled into the PPWG by a taper. EM simulations (Figure fig:char(a)) [Fallahi2014] were utilized to optimize the taper and calculate the coupling efficiency. Inside, a copper film photocathode serves as the bottom plate of the PPWG. There, a UV pulse (estimated duration $\tau_{uv} = 275$ fs) backilluminates the film, producing electrons inside the PPWG by photoemission. Concurrently, the THz field accelerates the electrons vertically across the PPWG. The electrons exit the gun through a slit on the top plate (anode) and are spectrally characterized by a retarding field analyzer (RFA) or counted by a Faraday cup. Figure fig:char(b) shows the THz-accelerated charge versus UV position along the slit. Both UV and THz pulses are generated from the same laser, ensuring absolute timing synchronization.

The THz pulse is focused into the gun with a maximum impinging energy of 35.7 µJ. EO sampling at PPWG-center (location of the center of the gun with the gun removed) and PPWG-thru (focus of a image-relay following propagation through the PPWG) reveals single-cycle durations of $\tau_{THz} = 1.2$ ps (Figure fig:char(d)), confirming that the PPWG induces minimal dispersion. Taking into account the energy, waveform, beam profile (Figure fig:char(f)), and coupling efficiency, the THz pulse has a calculated peak field of 153 MV/m in freespace and 350 MV/m in the PPWG.

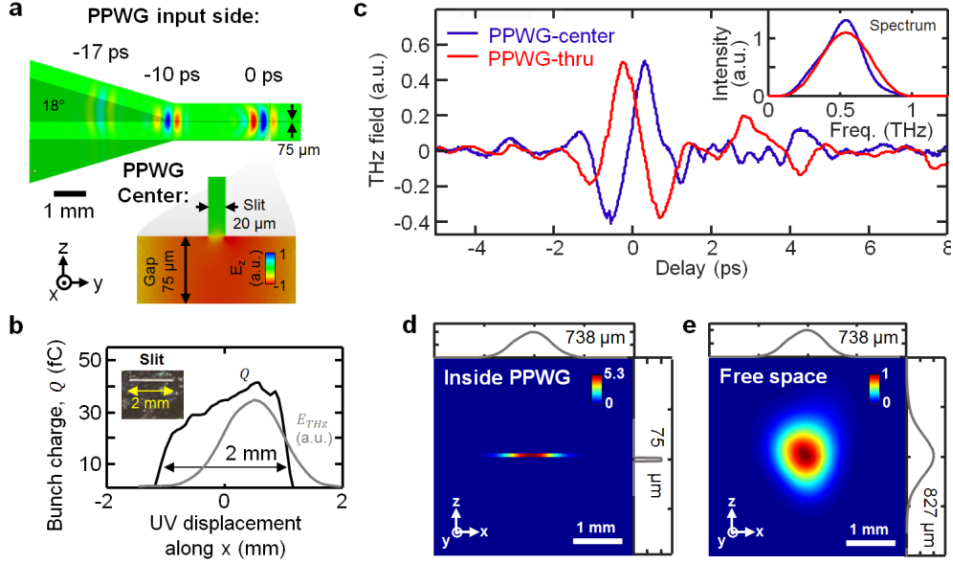

**Figure fig:char | Characterization of the gun. (a)** Several snapshots of the THz wave coupling into the PPWG, based on EM simulations. (Inset) Magnified cross-section shows that the field amplitude inside the gun is highly uniform and unperturbed by the slit. **(b)** Total THz-accelerated bunch charge exiting the gun as a function of the UV photoemitter displacement along the direction of the slit (black), creating a 1D electron emission map in the presence of the spatially-varying focused THz field (gray). **(c)** Temporal profiles measured via EO sampling of the THz electric field at PPWG-center and PPWG-thru (scaled). (See main text for definitions.) **(d)-(e)** THz beam intensity **(d)** inside the PPWG [calculated from **(e)**] and **(e)** at the free-space focus (measured). The colorbars of the two beam profiles show a 5.3x intensity enhancement in the PPWG.

## 3. Results and discussion

The electron momentum gain, $p_e$, can be expressed as $p_e(t_{emit}) = q \int_{t_{emit}}^{t_{escape}} E_{thz}(t) \mathrm{d}t \approx q A_{THz}(t_{emit})$, where $A_{THz}$ is the THz vector potential, $t_{emit}$ is the emission time (i.e. delay), and $t_{escape}$ is the time the electron exits the PPWG. The approximation $p_e(t_{emit}) \approx q A_{THz}(t_{emit})$ is valid because $t_{escape} \gg \tau_{THz}$ in our setup (shown later). To determine the optimum emission time for acceleration, we record the electron energy gain ($W_e$) spectra and bunch charge versus delay in Figures fig:specgram(a)-(b). The UV emitter can precede (<2 ps), overlap (-2 to 2 ps), or succeed (>2 ps) the THz pulse. In the overlap region (-2 to 2 ps), $W_e$ maps out the phase and amplitude of $A_{THz}(t_{emit})$, similar to THz streaking in gases [Fruhling2009]. One exception is that between -0.25 and 0.4 ps, emission occurs in the positive half-cycle of the THz field, causing a suppression of charge and energy gain. Two delays are selected to be the operating points of the gun. The first delay, $\tau_1 = -2$ ps, produced the highest peak acceleration while the second delay, $\tau_2 = 0.8$ ps, produced the most monoenergetic spectra. The total bunch charge was 40 fC at $\tau_1$ and 32 fC at $\tau_2$.

When the emission precedes the THz pulse (<2 ps), a large energy spread centered about ~0.45 keV is observed. The origin of these broadened spectra, enduring for long decay times, is attributed to multiple complex mechanisms encompassing thermal [Herink2014] or time-of-flight effects. Further discussion is provided in Supplementary Information. When the emission succeeds the THz pulse (>2 ps), there is no net acceleration from that pulse. The constituency of electrons slightly elevated to 50 eV is attributed to the aforementioned decay effects probed by a backreflected THz pulse arriving at 18 ps (Supplementary Information).

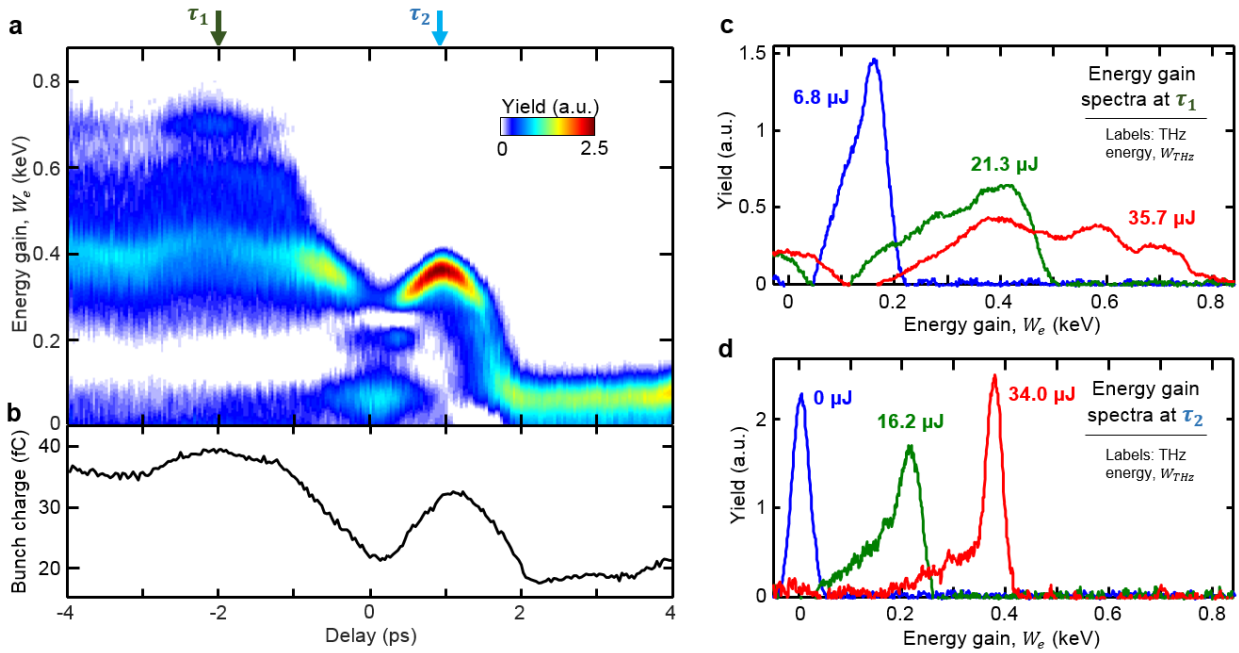

**Figure fig:specgram | THz-driven electron energy gain and bunch charge modulation. (a) Measured spectrogram showing the energy gain spectra as a function of delay between UV and THz pulses, at maximum THz energy. (b) Measured bunch charge as a function of delay. (c)-(d) Electron energy spectra for three different THz energies at delay position (c) $\tau_1 = -2$ ps and (d) $\tau_2 = 0.8$ ps.**

In Figures fig:specgram(c)-(d), we take a closer look at the energy spectra from the two operating points, $\tau_1$ and $\tau_2$, for three different THz energies, $W_{THz}$. Each spectrum exhibits a unimodal distribution with an average energy gain increasing with $W_{THz}$. Except for the $W_{THz} = 35.7$ μJ spectrum at $\tau_1$, the spectral shapes are asymmetric with a pedestal toward lower energies and a maximum yield toward higher energies, followed by a sharp cutoff, akin to the shapes observed in RF accelerators [Warren1983]. The high yield near the cutoff indicates that most electrons are emitted at the optimal THz phase and concurrently experience the same acceleration. The pedestal can be attributed to electrons emitted away from the optimal phase, resulting in a lower energy gain.

We continue investigations at $\tau_1$ and $\tau_2$ by plotting $W_e$ versus $W_{THz}$ on a spectrogram (Figures fig:scaling(a)-(b)) and scatter plot (Figures fig:scaling(c)-(d)). At both delays, $W_e$ scales mostly linearly with $W_{THz}$ or, equivalently, with $E_{THz}^2$. This scaling law can be explained by $W_e = p_e^2/2m \propto E_{THz}^2$, which is valid when $t_{escape} \gg \tau_{THz}$. Alternatively, if $t_{escape} \ll \tau_{THz}$, the energy gain would be dominated by $W_e = q \int_{z_{emit}}^{z_{escape}} E_{THz}(z) \mathrm{d}z$, leading to a $W_e \propto E_{THz}$ scaling law, as is typical in RF guns [Harris2011] and would be the case in this study for larger field or reduced PPWG spacing.

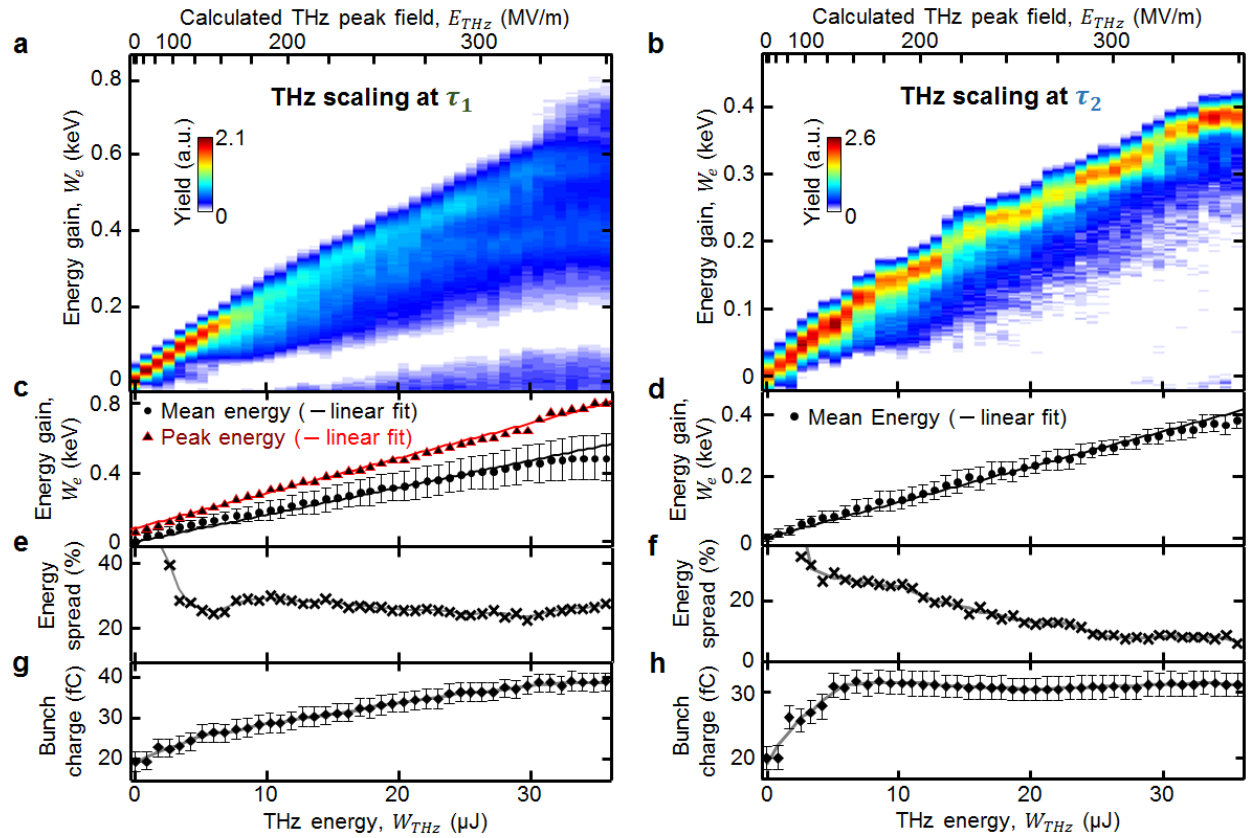

Figure fig:scaling | THz scaling at $\tau_1$ and $\tau_2$, delay positions defined earlier in Figure fig:specgram(a). (a)-(d) Energy gain plotted on a spectrogram and scatter plot to highlight its scaling as a function of accelerating THz energy or THz field. Error bar radius is equal to the absolute RMS energy spread. (e)-(f) Percent RMS spread of the accelerated bunch. (g)-(h) Total detected bunch charge exiting the gun. Error bar radius is equal to the RMS instrument noise.

At $\tau_1$, increasing the THz energy results in an increase of absolute energy spread (Figure fig:scaling(a)). Consequently, the relative energy spread remains roughly constant at around 20-30% (Figure fig:scaling(e)). The bunch charge increases monotonically with THz energy (Figure fig:scaling(g)). We obtain a peak energy gain of 0.8 keV at $W_{THz} = 35.7$ µJ (Figure fig:scaling(c)).

At $\tau_2$, the absolute energy spread remains constant with THz energy (Figure fig:scaling(b)). Correspondingly, the percent energy spread monotonically decreases with THz energy, to a minimum of 5.8% centered near 0.4 keV (Figure fig:scaling(f)). The pedestal regions are neglected in the energy spread calculations, since over time those electrons separate from the main bunch. Half of this spread comes from THz shot-to-shot fluctuations (2%), while another large contribution comes from the spread in electron emission time: $\Delta t_{emit} = \tau_{uv} = 275$ fs $= T_{THz}/8$. By stabilizing the laser and shortening $\tau_{uv}$ via an OPA [Ziegler1998], the energy spread can be further reduced. In Figure fig:scaling(h), the bunch charge increases with THz energy below 7 µJ, indicating that the emission is space-charge-limited [Rosenzweig1994]. Above 7 µJ, the bunch charge plateaus, indicating that the THz field overcomes the space charge force and extracts all the emitted electrons.

## 4. Simulations

In Figure fig:sim(b), we show the calculated single electron energy gain versus delay using the measured THz waveform with a fitted field strength (Figure fig:sim(a)), overlaid with the measured peak energy gain from Figure fig:specgram(a). Several experimental features are represented in this simple analytical model: (1) suppression region around 0 ps, (2) relative energy gain levels and (3) delay between the two peaks. This model also provides an alternate method for quantifying the THz field strength inside the gun. Our fitted peak field was 480 MV/m.

To better understand the bunch dynamics under the influence of space charge and THz field, particle tracking simulations in Figures fig:sim(d) and (f) show the evolution of the (d) energy spectrum and (f) temporal profile of the 32 fC bunch emitted at $\tau_2$ as it propagates along z. The THz pulse is passed by the time the bunch reaches 25 µm, verifying $t_{escape} \gg \tau_{THz}$. At the gun exit ($z = 75$ $\mu m$), the simulated energy spectrum has excellent overlap with the experimental spectrum (Figure fig:sim(c)). The sharp cutoff, pedestal height, pedestal length, and central lobe width are all reproduced flawlessly by the model. The simulated temporal profile at the gun exit (Figure fig:sim(e)) exhibits a pulse duration of 321 fs, longer than the initial 275 fs due to space charge.

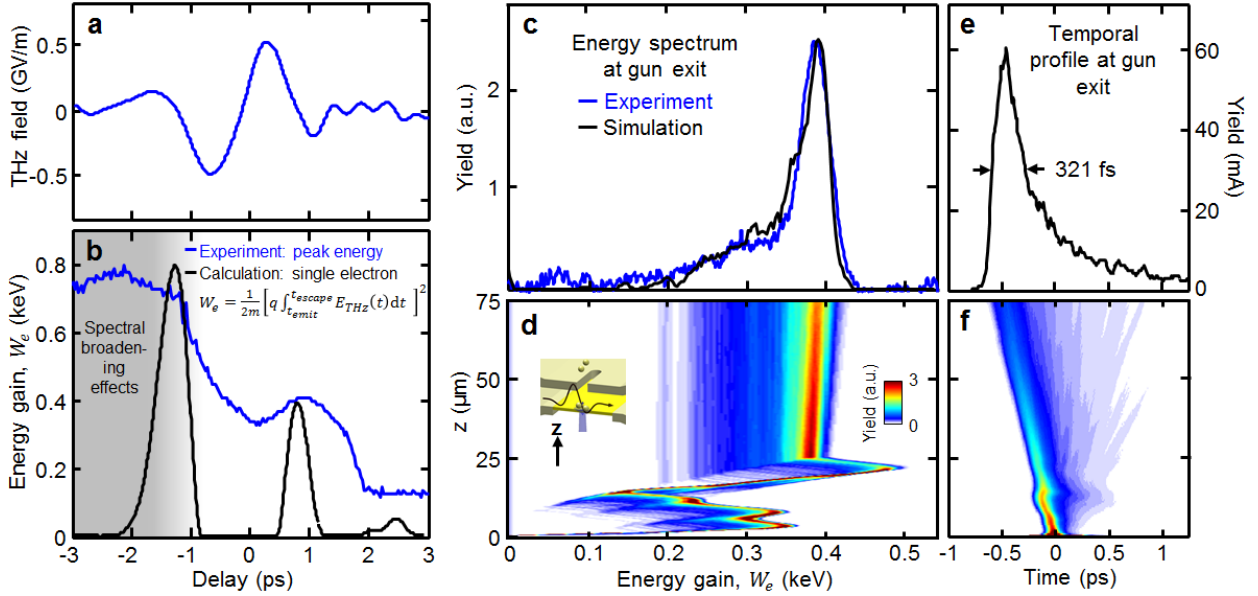

**Figure fig:sim | Numerical analysis of THz gun. (a)** THz electric field measured by EO sampling with fitted field strength. **(b)** The single electron energy gain, calculated analytically, is overlaid with the peak energy gain obtained from experiment in Figure fig:specgram(a). Using the experimental peak energy gain as a comparison is justified because it represents the gain of a single electron emitted at the optimal delay and spatial position. **(c)** Simulated energy spectrum of the bunch at the gun exit for emission at $\tau_2$, showing excellent agreement with experiment. **(d)** Simulated evolution of the energy spectrum along z. The THz pulse is passed by the time the electrons reach 25 μm. **(e)** Temporal profile of electron bunch at the gun exit, showing a FWHM pulse duration of 321 fs, elongated by space charge. **(f)** Simulated evolution of the temporal profile along z. Time zero is defined to be the centroid of the bunch in the frame of the moving bunch. These simulations utilized a particle tracking code which incorporated space charge, imitated the experimental conditions, and used the THz field profile in (a).

## 5. Conclusion

In conclusion, we demonstrated high field (>300 MV/m), quasimonoenergetic (few percent spread) THz acceleration of multi-10 fC electron bunches to sub-keV energies in an ultracompact, robust device. No degradation in performance was observed over 1 billion shots. While the operating pressure was 40 μTorr, no change in performance was observable up to 10 mTorr. This first result of a jitter-free, all-optical THz gun, powered by a few-mJ laser, performs in accordance with underlying simulations and is encouraging for future developments. In its current state, it can be used for time-resolved LEED and—with modest improvements in laser stability and $\tau_{uv}$—for time-resolved electron energy-loss spectroscopy (EELS) [Piazza2014]. Further improvements on the gun structure and THz field promise relativistic electrons [Fallahi2016].

## 6. Supplementary Information

### 6.1. Delay scan

In order to have a fuller understanding of the electron dynamics induced inside the gun by the THz and UV pulses, we acquire a spectrogram over a wide range of delays in Figure fig:longdelay(a). Between -2 and 2 ps, the electron spectra change rapidly with respect to delay due to the temporal overlap with the main THz pulse. Here, the spectra are narrowband and the momentum gain follows the vector potential of the THz field, as described in the main text.

When the UV pulse precedes the THz pulse (<-2 ps), we observe broad, elevated electron spectra enduring over a long delay window to nearly -50 ps. A number of physical processes may contribute to this behavior. Detailed investigations will be the topic of a forthcoming article.

One possibility is thermally-assisted THz field emission, a process investigated in [Herink2014] and more generally in [Fujimoto1984,Hohlfield2000]. As UV photons are absorbed, electrons are promoted in energy. The increase in kinetic energy causes an elevated electron temperature distribution (i.e., hot electrons) over a period of tens of fs. The hot electrons then collide with phonons to dissipate heat to the lattice via electron-phonon collisions, resulting in an elevated lattice temperature, or a smeared-out Fermi-Dirac distribution with a high energy tail. This elevated lattice temperature decays over a ps time scale, as is known from experiments on similar thin films [Hohlfield2000]. When the THz field impinges the surface during this time and lowers the Schottky barrier, electrons in the higher tail of the distribution have an increased tunnelling probability. Once emitted, the electrons are subject to freespace THz acceleration. Unlike UV photoemission, which creates an electron bunch of defined duration, the field-emitted electrons here can be emitted over a wide range of THz phase, so long as the field can sufficiently lower the Schottky barrier to enable tunnelling. Consequently, as the THz field increases, the spectra grow broader (Figure fig:scaling(a)) and the emitted charge increases (Figure fig:scaling(g)).

Figure fig:longdelay(b) shows the normalized current as a function of delay. We observe exponential decay behavior preceding the main (<0 ps) and backreflected (2 to 18 ps) THz pulses. The base level of current is about a factor of 0.48 times the maximum current. A decay time can be determined by fitting an exponential function (offset by the base level) to the normalized current, shown as a red/blue curve for the decay preceding the main/backreflected THz pulse. We find that the exponential decay time is 16.7 ps for both curves, suggesting that there is an underlying thermal relaxation constant that is independent of THz

field strength. This decay time is comparable to that measured from transient reflectivity measurements on similar thin metal films in [Hohlfield2000], which was, e.g., ~10 ps for a 20 nm Au film at 1 mJ/cm² pump fluence.

Another possibility for the decay behavior is time-of-flight effects. The bias voltage of 9 V in the 75 μm gap between the two plates of the PPWG implies that an electron released on the cathode takes 84 ps to reach the anode ($t_{escape} = 84$ ps). During this time, the bunch can be manipulated by the arriving THz pulse.

When the UV pulse succeeds the THz pulse (>2 ps), we observe a scaled-down (in energy) replica of the aforementioned decay effects. This can be attributed to the presence of a weak, backreflected THz pulse arriving later at 18 ps (by definition, the envelope of the main THz pulse arrives at 0 ps). Physically, the backreflection occurs in the gun at the interface between the end of the PPWG and the output taper, which is 2.75 mm of propagation away from the exit anode (see inset in Figure fig:decay(a)). The roundtrip propagation of 5.5 mm matches well with the arrival of the backreflected THz at 18 ps.

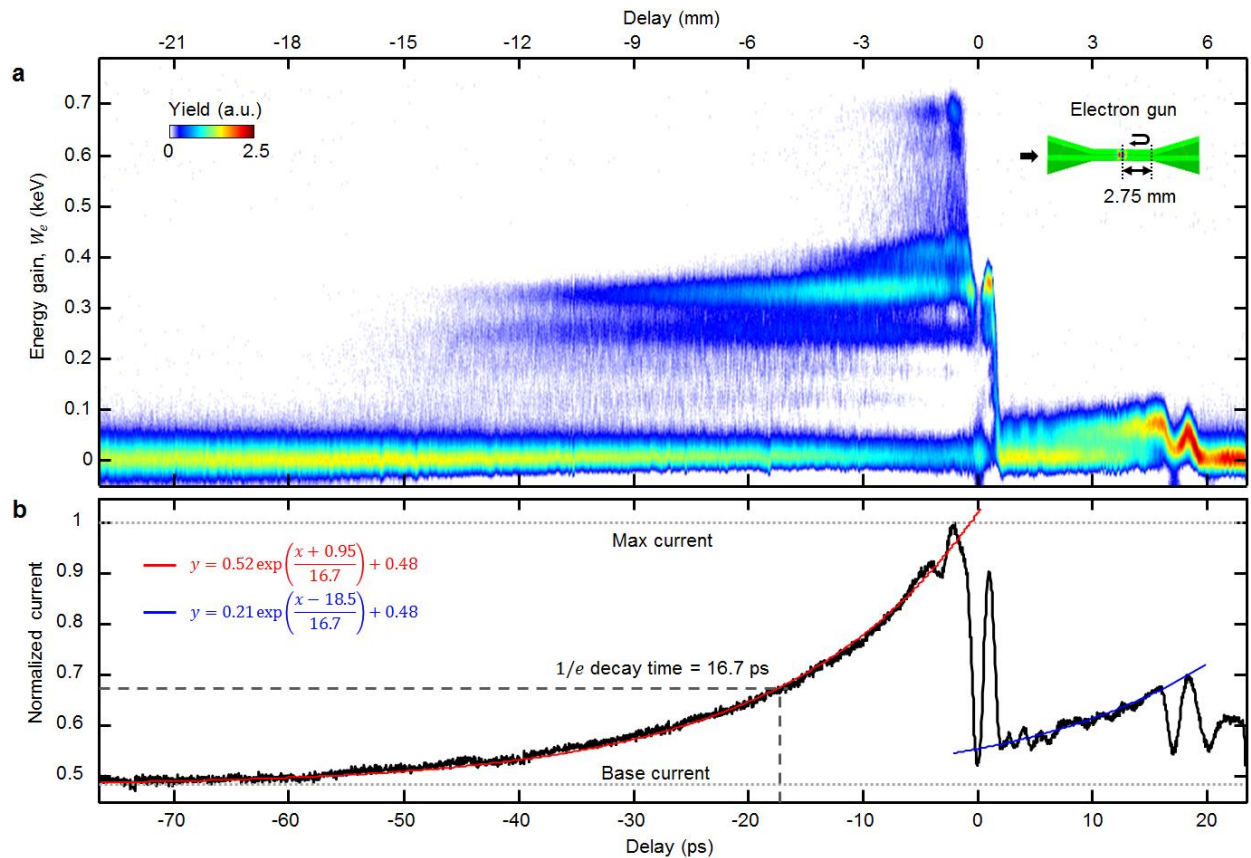

**Figure fig:decay | Delay scan. (a) Spectrogram showing the energy gain spectra over a wide range of delays. Broad electron spectra is exhibited for UV pulse delays of up to 50 ps preceding the THz pulse. The presence of weak, backreflected THz pulses is evident at 18 ps. (Inset) This backreflection occurs at the interface between the PPWG and the output taper, which is 2.75 mm away from the exit anode. (b) Normalized current as a function of delay, showing exponential decay behavior preceding the main (0 ps) and backreflected (18 ps) THz pulses. Exponential curve fitting determines the decay time to be 16.7 ps.**

## 6.2. THz spatiotemporal properties as a function of energy

In Figure fig:scaling, the THz energy, $W_{THz}$, was varied by changing the IR pump energy and measured using a pyroelectric detector. Since the acceleration process depends on the spatiotemporal properties of the THz beam, it is important to verify that there are no significant distortions in the THz temporal and spatial profiles as the energy is changed. It is also important to verify that the THz field strength scales proportionally as the square root of the THz energy.

We first measure the temporal profile via EO sampling for various THz energies in Figure fig:spatiotemporal(a). Aside from scaling in field strength, the shape of the temporal profile has little variation as a function of THz energy, with the carrier-envelope phase and pulse duration remaining roughly constant. In Figure fig:spatiotemporal(b), we sample the THz field as a function of THz energy at two peaks of the waveform as labelled in Figure fig:spatiotemporal(a): Peak 1 (black dots) and Peak 2 (gray dots). The field strengths at these two peaks correspond to the maximum accelerating fields experienced by the electron in the gun. The data fits well to a square root function (dashed lines), thus verifying that the peak accelerating field scales with the square root of the THz energy.

Next we measure the spatial profile of the THz beam at the freespace focus using a THz camera for various energies in Figure fig:spatiotemporal(c)-(e). The normalized horizontal and vertical lineout profiles are overlapped in Figure fig:spatiotemporal(c)-(d), revealing negligible variation in their Gaussian-like shapes. Further, the $1/e^2$ beam diameters are plotted in Figure fig:spatiotemporal(e) as a function of THz energy for the vertical (black circles) and horizontal (gray squares) profiles. The diameters vary by an average of only 3.7% (horizontal) and 2.9% (vertical) with respect to the mean (dotted lines) over the range of THz energies. The 2D beam profiles for several THz energies are shown in the inset, revealing negligible variation.

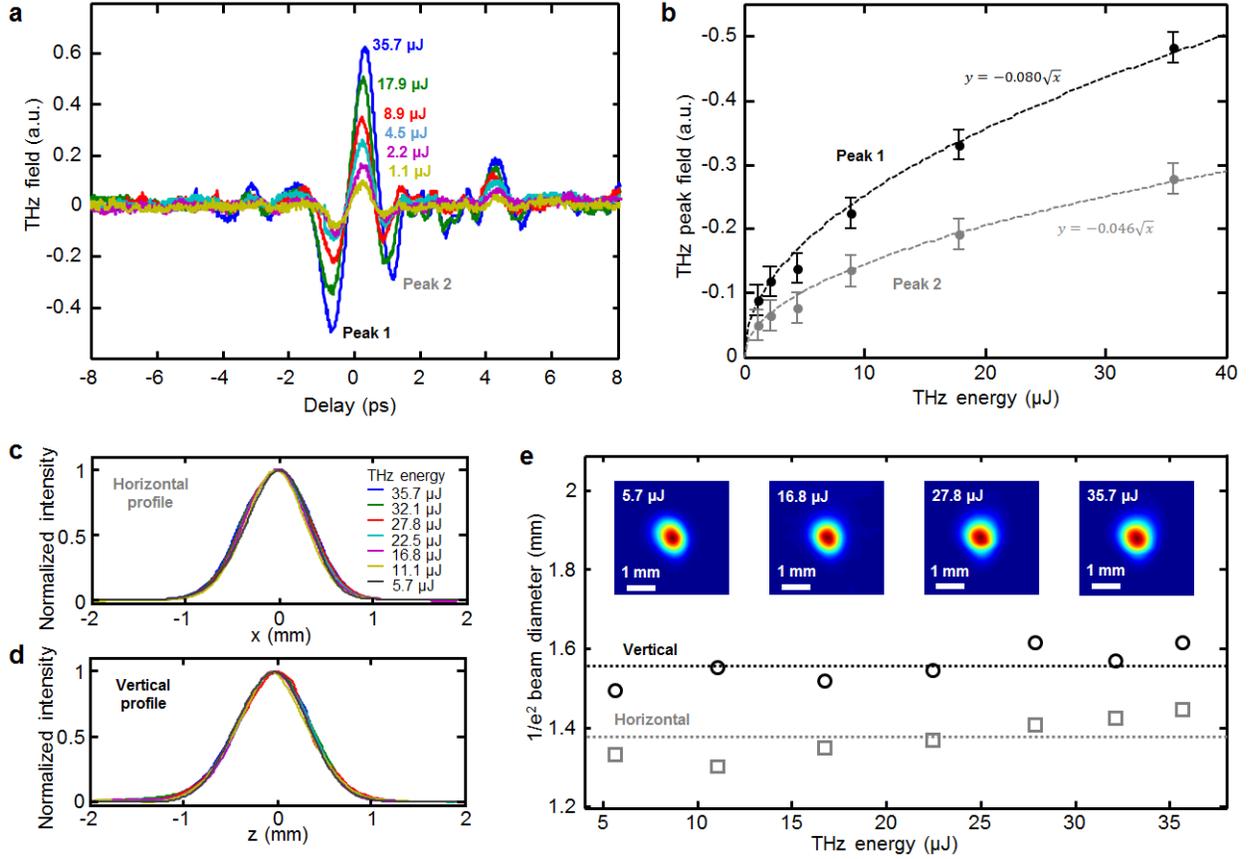

**Figure fig:spatiotemporal | THz spatiotemporal properties as a function of energy. (a)** EO sampling waveforms for various THz energies. **(b)** The relative field strength as a function of THz energy at Peak 1 (black dots) and Peak 2 (gray dots) of the waveform, as labelled in **(a)**. The measured peak field strengths scale as the square root of the THz energy according to the fits (dashed lines). The error bar radius is determined by calculating the RMS noise of the EOS signal in the absence of THz. **(c)** The normalized horizontal and **(d)** vertical lineout THz beam profiles at the freespace focus for various THz energies. Profiles are Gaussian with minimal variations in shape as a function of THz energy. **(e)** The horizontal (gray squares) and vertical (black circles) $1/e^2$ beam diameters as a function of THz energy, shown here to all reside near their respective collective mean value (dotted lines). **(Inset)** 2D beam profiles at several sample THz energies, showing minimal variation.

### 6.3. Calculation of coupling efficiency into the gun

Here, we show how the THz coupling efficiency into the gun, $\eta_{gun}$, can be calculated from the measured power transmission data, $T$, shown in Figure fig:vna (blue line).

First, we make the assumption that the out-coupled freespace mode, denoted by $E_{ppwg}(x, y, z)$ and shown in Figure fig:ccalc(b), varies minimally for different PPWG spacings, $d$. This has been validated by EM simulations for values of $d$ within our region of interest: $0 < d < 200\ \mu m$. This mode is the beam which couples most efficiently from freespace into the TEM mode of the PPWG, with a coupling efficiency denoted by $\eta$. Our THz beam in-coupled into the PPWG can be approximated as a fundamental Gaussian beam, denoted by $E_{gaussian}(x, y, z)$ and also shown in Figure fig:ccalc(b). The amount power in the $E_{ppwg}(x, y, z)$ component of the $E_{gaussian}(x, y, z)$ mode, as a fraction of the total power, is denoted by $F$. Using EM simulations, we determine $F$ by computing the overlap integral over a chosen plane normal to y:

$$F = \frac{\left| \int E_{gaussian}^{*} E_{ppwg} \mathrm{d}x\mathrm{d}z \right|^{2}}{\int \left| E_{gaussian} \right|^{2} \mathrm{d}x\mathrm{d}z \int \left| E_{ppwg} \right|^{2} \mathrm{d}x\mathrm{d}z}$$

Note the integrand is scalar because the modes have only one and the same polarization. We obtained a result of $F = 0.8$.

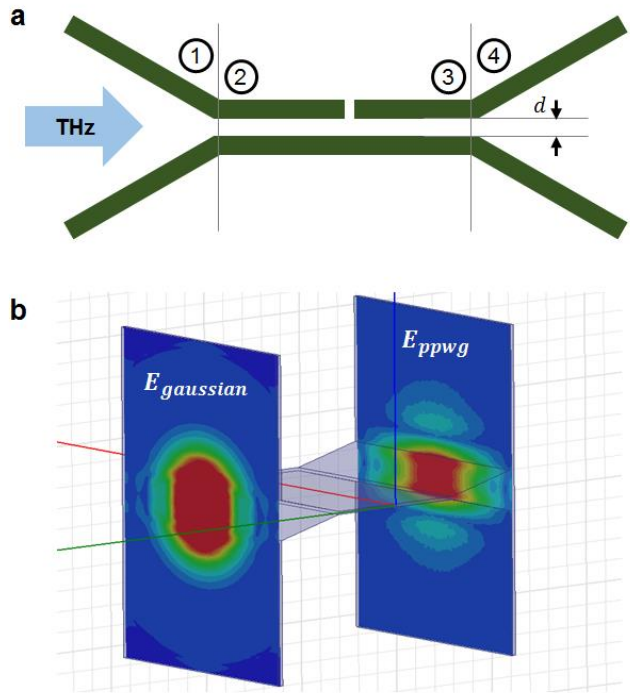

**Figure fig:ccalc | Calculation of THz coupling efficiency into the PPWG gun. (a)** Schematic of the **PPWG gun denoting the interfaces at which reflections occur. (b)** EM simulations showing the in-

coupled freespace Gaussian mode ($E_{gaussian}$) and the out-coupled freespace mode of the TEM waveguide ($E_{ppwg}$).

Second, we assume that the large majority of transmission losses come from wall ohmic losses inside the PPWG (region 2↔3 in Figure fig:ccalc(a)) and from reflections at the interfaces between the PPWG and taper sections (regions 1↔2 or 3↔4 in Figure fig:ccalc(a)). We proceed to express the power transmission and reflection at the interfaces as follows.

$$\text{Transmission } 1 \to 2: F\eta$$
$$\text{Transmission } 3 \to 4: \eta$$
$$\text{Reflection } 3 \to 4: 1 - \eta$$

A reflected wave at the 3→4 interface propagates backward toward the 2→1 interface. There it experiences a second reflection.

$$\text{Reflection } 2 \to 1: 1 - \eta$$

Also, the propagation along 2↔3 induces ohmic losses. The propagation efficiency of one pass is denoted by

$$\text{Propagation } 2 \leftrightarrow 3: \beta$$

Using EM simulations with finite conductivity surfaces, we determined $\beta = 0.83$. We can now calculate the total transmission through the structure:

$$\text{Transmission } 1 \to 4:$$
$$T = F\eta[\eta + \beta^2(1-\eta)^2\eta + \beta^4(1-\eta)^4\eta + \cdots]$$

After some algebraic simplification, we obtain

$$T = \frac{F\eta^2}{1 - \beta^2(1-\eta)^2}$$

This equation gives the power transmission through the waveguide, $T$, as a function of the TEM mode coupling efficiency, $\eta$. Since we have measurements of $T$ and wish to know $\eta$, we reverse the equation:

$$\eta = \frac{\beta^2 T + \sqrt{[F - \beta^2(F + T)]T}}{F + \beta^2 T}$$

Finally, the coupling efficiency of our THz beam from freespace into the center of the gun, $\eta_{gun}$, as a function of $\eta$ is simply

$$\eta_{gun} = F\sqrt{\beta}\,\eta$$

## 7. Methods

### 7.1. Pump laser

We used a 1 kHz, 1030 nm Yb:KYW regenerative amplifier [Calendron2014] seeded with a 42.5 MHz Yb:KYW oscillator from Amplitude Systemes. The pulses are compressed to the transform limit of 550 fs ($sech^2$) FWHM with shot-to-shot energy fluctuations of 0.5%. The available 4.2 mJ compressed pulses are split: 99% for THz generation and 1% for UV generation. For THz generation, the impinging beam onto the lithium niobate crystal has an energy of 3.4 mJ and a $1/e^2$ diameter of 2.0 mm (sagittal) and 3.4 mm (tangential).

### 7.2. THz source

We employed the tilted pulse front (TPF) pumping technique in a 5.6% MgO-doped congruent lithium niobate (LN) crystal [Hebling2002] cooled to 100 K. The IR pump beam is diffracted off a 1500 l/mm grating to acquire a TPF, which is then subsequently imaged—in the tangential direction—onto the LN using a 150 mm cylindrical lens. Another cylindrical lens with a sagittal focal length of 100 mm was used to shape the impinging pump beam for highest efficiency. The extracted optical-to-THz energy conversion efficiency was near 1.0% with 35.7 μJ of THz energy. We used a Gentec SDX-1152 calibrated pyroelectric THz joulemeter to measure the THz energy. Concurrently, a thermal power meter (Ophir Optronics) measured 18 mW at 1 kHz. A Spiricon Pyrocam IV camera was used to image the THz beam. Shot-to-shot energy fluctuation was 2%.

### 7.3. EO sampling

We employed an oscillator-based EO sampling setup since the pulses from the amplifier were too long to effectively probe the THz waveform. Oscillator probe pulses were overlapped with THz pulses on a 200 μm, 110-cut ZnTe crystal. The probe pulses sample the THz-induced birefringence as a function of delay and are subsequently interrogated by a quarter-wave plate, polarizer, and photodiode combination, as typical [Wu1995]. Because of the much higher repetition rate of the oscillator pulse train, a boxcar averager (SRS SR250) was used to electronically gate out the pulse that overlapped with the THz. The EO crystal mount was custom fabricated such that the crystal is in the same position as the center of the gun for the PPWG-center measurement in Figure fig:char(c). For the PPWG-thru measurement, the gun was placed in its operating position and the transmitted THz beam was image-relayed via two additional parabolic mirrors onto a second focus, where the EO crystal was then placed.

### 7.4. UV source

The UV photoinjection was obtained by frequency-quadrupling the fundamental 1030 nm pump. The first second-harmonic generation (SHG) stage consisted of a 0.5 mm thick type I BBO crystal with φ=23.7°, generating approximately 3 μJ of 515 nm pulses. The second SHG stage consisted of 0.5 mm thick type I BBO crystal with φ=44.6°, generating 600 nJ of 258 nm pulses. The conversion efficiency from fundamental to UV was about 7%. A CaF$_2$ prism was used to spatially separate the various wavelengths. The prism-induced dispersion over the subsequent 0.5 m propagation was determined through calculations to cause negligible increase of the pulse duration. The UV energy impinging the copper photocathode was 270 nJ. Both nonlinear conversions were in the unsaturated regime and the phase-matching bandwidths of the two BBO crystals are broader than the spectral bandwidths of both the 1030 nm and 515 nm pulses. Therefore, we estimate of the UV pulse duration as roughly half that of the fundamental. The focused UV beamwaists on the photocathode were 20 μm (x) and 60 μm (y).

## 7.5. THz gun design and characterization

A variety of THz gun structures were proposed, including rectangular waveguides, pillbox structures, and multi-cell standing wave structures. For the first demonstration, we opted to use a simple parallel plate waveguide (PPWG) structure because of its simplicity and compatibility with broadband THz pulses. The PPWG, having a subwavelength spacing of $d = 75$ μm, guides only the TEM (TM$_0$) mode. This mode has zero cutoff frequency and a propagation constant given by $k_z = \omega/c$ for all frequencies regardless of the spacing between the plates [Kong2000], and we leverage this property for broadband, unchirped enhancement of the THz field [Iwaszczuk2012].

Two parallel plates, fashioned with 18° tapers, were fabricated separately and afterwards sandwiched together with high-precision Kapton shims in-between to set the spacing and enforce parallelicity. We ultimately operated with a shim thickness of 75±15 μm after optimization (see next paragraph). EM simulations were performed in HFSS to obtain the optimal taper angle for efficient coupling. Fabrication of the THz waveguide was performed in-house using conventional machining tools. A flatness tolerance of 5 μm over a 1 in$^2$ area was specified for the parallel plate sections. A 9V reverse bias was applied across the plates to help with electron extraction.

A THz network analyzer was used to characterize the power transmission, $T$, through the PPWG for various spacings, as shown in Figure fig:vna. EM calculations helped to determine the power coupling efficiency into the gun, $\eta_{gun}$ (light blue), as a function of $T$ (blue) (Supplementary Information). Although $\eta_{gun}$ increases with spacing $d$, the field strength inside the gun, $E_{THz}$, is a trade-off between coupling efficiency on one hand, and field confinement on the other, as expressed by $E_{THz} \propto \sqrt{\eta_{gun}/d}$. It is desirable to have the highest field strength possible inside the gun. By plotting the normalized field strength, $E_{THz}$ (green), as a function of spacing, we found the optimal value at a spacing of 75 μm and a coupling efficiency of 0.3.

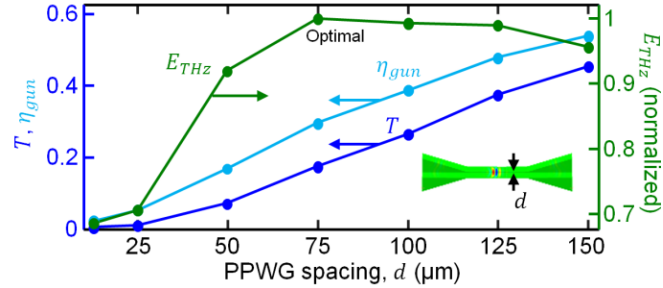

**Figure fig:vna | Measurement of the PPWG power transmission, $T$ (blue), and the corresponding power coupling efficiency into the gun, $\eta_{gun}$ (light blue), for various PPWG spacings. Based on these values, the normalized field inside the gun, $E_{THz}$ (green), can be determined. A spacing of 75 μm optimizes the normalized field strength.**

### 7.6. Photocathode

A 25 nm copper film coated on a UV-grade quartz substrate was used as the photocathode. The coating of the photocathode was performed in-house by evaporative physical vapor deposition. A chromium adhesion layer of a few nm was first deposited onto the substrate. In addition to functioning as a photocathode, the copper film functions as one of the PPWG plates. To minimize interface losses, the surface of the photocathode is placed flush with the surface of the parallel-plate structure to a tolerance of a few microns using an optical-flat mirror surface. To minimize THz diffraction losses through the thin film, the film is thickened to 125 nm (equal to the copper skin depth at 0.5 THz) along the THz propagation path from the PPWG input until ~0.25 mm before the 25 nm thick photoemission region. To ensure electrical connectivity between the photocathode and the PPWG, the copper film extended around to the edges of the quartz substrate and the edges were in contact with the PPWG.

### 7.7. Exit anode

The exit anode was cut out of a slab of 100 μm polished stainless steel shim stock. The precise slit width of 20 μm over a 2 mm length was achieved by picosecond laser micromachining. An optical microscope was used to verify the dimensions to within a tolerance of 2 μm. EM simulations in Figure fig:char(a) confirm that, with a width of 20 μm ($\lambda_{THz}/33$), the slit causes minimal distortion to the THz field distribution.

## 7.8. Electron detection

Following its exit from the gun, the electron bunch drifts into a retarding field analyzer (RFA) [Brunner2013], consisting of a channel electron multiplier (CEM) (Photonis, Inc.) and two static, uniform field regions formed by two biased mesh electrodes. The first region (between the gun and first electrode) boosts the electron energy by 300 eV to enhance the detection efficiency of the CEM. The second region (between first and second electrodes) acts as a highpass filter for the electron energy by retarding the electron trajectory using a variable bias $-V_{bias}$. Electrons having energy less than $eV_{bias}$ are repelled by the electrode while those with more energy pass through, being subsequently detected by the CEM. Each spectrum was collected by taking the derivative of the measured current with respect to $V_{bias}$. The intrinsic energy resolution of the analyzer is about 2 eV. With post-process smoothing, the effective resolution is about 16 eV.

The meshes were TEM grids (Ted Pella, Inc.) with a thickness of 13 µm and a pitch of 12.5 µm. Each mesh had a transmission of 36%, so the total bunch charge was determined from dividing the detected charge by $0.36^2 = 0.13$. Each mesh was placed on 100 µm thick stainless steel soldering plates and sandwiched by 500 µm PEEK shims which enforce their spacing and parallelicity as well as providing electrical isolation. The soldering plates contained "fingers" on which high voltage biasing wires were soldered. The RFA was placed 1.5 mm from the exit anode of the electron gun, measured by the distance between their nearest planes.

Absolute charge measurements were obtained by rewiring the grounded input terminal of the CEM to a Keithley 6514 picoammeter and turning off the CEM bias. In this configuration the CEM essentially acts as a Faraday cup. Secondary electron emission on the CEM is not taken into account, but it would only increase the total charge count if it were. The picoammeter had a RMS noise level of 300 fA.

## 7.9. Network analyzer measurements

Our vector network analyzer (VNA) setup for characterizing the PPWG power transmission consisted of an Agilent E8363B and millimeter wave extender V03VNA2-T/R with 70 dB of dynamic range at 0.220-0.325 THz. The transmitter and receiver were connected to corrugated horns designed for coupling the VNA waveguide mode to a free-space Gaussian mode with a waist of 6 mm. We placed two THz polyethylene lenses with focal lengths of 25 mm a distance of 2f apart between the transmitter and receiver and set the background level. Given the waist of 6 mm (diameter of 12 mm) and focal length of 25 mm, the f-number is about 2, which is well-matched to the optimal f-number of our PPWG's taper section. For the VNA measurements we replaced the photocathode with a polished aluminium blocks to eliminate diffraction losses in the PPWG. The PPWG was placed in the center between the two THz lenses and the PPWG transmission was characterized. The PPWG spacing was varied by changing the thickness of the Kapton shims between the two plates (see Methods - THz gun design and characterization).

## 7.10. Particle tracking simulations

3D particle tracking simulations incorporating space charge were used to model the electron bunch evolution in the presence of the THz field. The emitted electron bunch had a Gaussian spatial profile with beamwaists of 20 µm (x) and 60 µm (y) and a Gaussian temporal profile with FWHM of 275 fs. The initial kinetic energy was 0.18 eV (equal to the excess energy) with a uniform momentum distribution over a half-sphere [Dowell2009]. 5000 macroparticles were used to represent a total bunch charge of 32 fC, corresponding to a charge of -40e per macroparticle. The trajectories were modeled by integrating the kinematic equations for every particle $i$ using a 4th order Runge-Kutta solver:

$$m\frac{d\boldsymbol{v}_i}{dt} = \boldsymbol{F}_{field} + \boldsymbol{F}_{bias} + \sum_j \left(\boldsymbol{F}_{image,ij} + \boldsymbol{F}_{coulomb,ij}\right)$$

$$\frac{d\boldsymbol{r}_i}{dt} = \boldsymbol{v}_i$$

Here, $m$ is the relativistic mass, $\boldsymbol{F}_{field}$ is the electric force due to the THz pulse, $\boldsymbol{F}_{bias}$ is the electric force due to the 9V reverse DC bias, $\boldsymbol{F}_{image,ij}$ is the force on the $i$th particle due to the $j$th image particle, and $\boldsymbol{F}_{coulomb,ij}$ is the particle-particle Coulomb force. The THz beam was modeled as a plane wave with a Gaussian distribution in amplitude in the x direction. The THz waveform in $\boldsymbol{F}_{field}$ was directly imported from the EO sampling trace.

## 9. Acknowledgements

We acknowledge Andrej Berg, Thomas Tilp, and the HasyLab (DESY) machine shop for mechanical fabrication, Stuart A. Hayes for advice regarding electron gun design, Wenjie Lu for initial photocathode coating runs, Johann Derksen for help with instrument troubleshooting, and Prof. Richard Temkin for allowing the use of his microwave measurement equipment. This work was supported by the European Research Council through Synergy Grant (AXSIS) 609920, by the Air Force Office of Scientific Research under grant AFOSR - A9550-12-1-0499, and by the excellence cluster "The Hamburg Centre for Ultrafast Imaging- Structure, Dynamics and Control of Matter at the Atomic Scale" of the Deutsche Forschungsgemeinschaft (by grant EXC 1074). W.R.H. acknowledges support by a National Defense Science and Engineering Graduate (NDSEG) Fellowship. X.W. acknowledges support from a Research Fellowship from the Alexander von Humboldt Foundation.

## 10. Author contributions

F.X.K. and A.F. conceived the concept of a THz-driven electron gun and acceleration using low energy THz pulses in the TEM mode of a PPWG, respectively. W.R.H. and A.F. designed the THz electron gun structure. Experiments were designed and carried out by W.R.H. with help from X.W. Simulations were performed by A.F. (DGTD/PIC) and W.R.H. (particle tracking and HFSS). W.R.H. and E.A.N. performed network analyzer based PPWG characterization. A.-L.C. and H.C. constructed and maintained the pump laser. W.R.H., A.-L.C., and K.-H.H. contributed to the UV generation. D.Z. coated the photocathode. K.R. contributed to the THz source optimization. W.R.H., A.F., K.R., E.A.N., and F.X.K. analyzed the data and interpreted the results. A.F. and F.X.K. provided feedback to improve the experiment. W.R.H. wrote the manuscript with revisions by all.

**Competing financial interests**: The authors declare no competing financial interests.

**Materials & Correspondence**: Correspondence and requests for materials should be addressed to F.X.K.